\documentclass[12pt]{article}
\usepackage{amsmath,amssymb}
\usepackage{epsfig} 

\def\m@th{\mathsurround=0pt }
\def\leftrightarrowfill{$\m@th \mathord\leftarrow \mkern-6mu
        \cleaders\hbox{$\mkern-2mu \mathord- \mkern-2mu$}\hfill
        \mkern-6mu \mathord\rightarrow$}

\def\overleftrightarrow#1{\vbox{\ialign{##\crcr
        \leftrightarrowfill\crcr\noalign{\kern-1pt\nointerlineskip}
        $\hfil\displaystyle{#1}\hfil$\crcr}}}

\def\simlt{\stackrel{<}{{}_\sim}}
\def\simgt{\stackrel{>}{{}_\sim}}

\newcommand{\be}{\begin{equation}}
\newcommand{\ee}{\end{equation}}

\def\shat{\ifmmode \hat{s}\else $\hat{s}$\fi}
\def\gp2{{g'}^2}
\def\g2{g^2}
\def\g32{g_s^2}

\newcommand{\newc}{\newcommand}

\newc{\gsim}{\lower.7ex\hbox{$\;\stackrel{\textstyle>}{\sim}\;$}}
\newc{\lsim}{\lower.7ex\hbox{$\;\stackrel{\textstyle<}{\sim}\;$}}
\newc{\ie}{{\it i.e.}}
\newc{\etal}{{\it et al.}}
\newc{\mev}{\hbox{\rm\,MeV}}
\newc{\gev}{\hbox{\rm\,GeV}}
\newc{\tev}{\hbox{\rm\,TeV}}
\newc{\xpb}{\hbox{\rm\, pb}}
\newc{\xfb}{\hbox{\rm\, fb}}

\newc{\G}{{\cal G}}
\newc{\h}{{\cal H}}
\newc{\D}{{\cal D}}
\newc{\E}{{\cal E}}

\newc{\mtop}{M_t}
\newc{\mbot}{m_b}
\newc{\mz}{M_Z}
\newc{\mw}{M_W}
\newc{\alphasmz}{\alpha_s(M_Z)}
\newc{\swsq}{\sin^2\theta_W}
\newc{\cwsq}{\cos^2\theta_W}
\newc{\tw}{\tan\theta_W}
\newc{\cw}{\cos\theta_W}
\newc{\sw}{\sin\theta_W}
\newc{\BR}{\hbox{\rm BR}}
\newc{\zbb}{Z\to b\bar}
\newc{\Gb}{\Gamma (Z\to b\bar b)}
\newc{\Gh}{\Gamma (Z\to \hbox{\rm hadrons})}
\newc{\sgn}{\mbox{sgn}}

\def\eq#1{eq.~(\ref{#1})}

\def\vev#1{\langle {#1} \rangle}

\newcounter{mysubequation}[equation]

\def\beq{\begin{equation}}
\def\eeq{\end{equation}}
\def\bea{\begin{eqnarray}}
\def\eea{\end{eqnarray}}
\def\slashchar#1{\setbox0=\hbox{$#1$}           
   \dimen0=\wd0                                 
   \setbox1=\hbox{/} \dimen1=\wd1               
   \ifdim\dimen0>\dimen1                        
      \rlap{\hbox to \dimen0{\hfil/\hfil}}      
      #1                                        
   \else                                        
      \rlap{\hbox to \dimen1{\hfil$#1$\hfil}}   
      /                                         
   \fi}                                         %
\catcode`@=11
\long\def\@caption#1[#2]#3{\par\addcontentsline{\csname
  ext@#1\endcsname}{#1}{\protect\numberline{\csname
  the#1\endcsname}{\ignorespaces #2}}\begingroup
    \small
    \@parboxrestore
    \@makecaption{\csname fnum@#1\endcsname}{\ignorespaces #3}\par
  \endgroup}
\catcode`@=12

\begin{document}

\baselineskip=18pt

\setcounter{footnote}{0}
\setcounter{figure}{0}
\setcounter{table}{0}

\begin{titlepage}
\begin{flushright}
CERN-PH-TH/2007--179\\
IFT-UAM/CSIC-07-50
\end{flushright}
\vspace{.3in}

\begin{center}
{\Large \bf Cosmological implications  of the\\ Higgs mass measurement}

\vspace{0.5cm}

{\bf J.R. Espinosa$^{a}$, G.F. Giudice$^{b}$} and {\bf A. Riotto$^{b,c}$}

\centerline{$^{a}${\it IFT-UAM/CSIC, 28049 Cantoblanco, Madrid, Spain}}
\centerline{$^{b}${\it CERN, Theory Division, CH--1211 Geneva 23, 
Switzerland}}
\centerline{$^{c}${\it INFN, Sezione di Padova, Via Marzolo 
8, I-35131 Padua, Italy}}

\end{center}
\vspace{.8cm}

\begin{abstract}
\medskip
\noindent
We assume the validity of the Standard Model  up to an arbitrary high-energy scale and discuss  what information on the early stages of the Universe can be extracted from a measurement of the Higgs mass. For $M_h \simlt 130 \gev$, the Higgs potential can develop an instability at large field values. From the absence of excessive thermal Higgs field fluctuations we derive a bound on the reheat temperature after inflation as a function of the Higgs and top masses. Then we discuss the interplay between the quantum Higgs fluctuations generated during the primordial stage of inflation and the cosmological perturbations, in the context of landscape scenarios in which the inflationary parameters scan. We show that, within the large-field models of inflation, it is highly improbable to obtain the observed cosmological perturbations in a Universe with a light Higgs. Moreover, independently of the inflationary model, the detection of primordial tensor perturbations through the $B$-mode of CMB polarization and the discovery of a light Higgs can simultaneously occur only with exponentially small probability, unless there is new physics beyond the Standard Model.

\end{abstract}

\bigskip
\bigskip

\end{titlepage}


\section{Introduction}

The search for the Higgs boson and the measurement of its properties are
one of the primary goals of the LHC. The mere discovery of a Higgs can be viewed as a possible
indication of additional new physics not far from the
electroweak scale, because of the high sensitivity to short-distance
quantum corrections of the mass term associated to a fundamental scalar.
However, even in the absence of any new-physics discovery at the LHC, a
measurement of the Higgs mass can give us useful hints on the structure of
the theory at very high energies. This is because, at large field values,
the Higgs potential can develop an instability or become
non-perturbative, depending on the precise value of the Higgs quartic
coupling $\lambda$ or, ultimately, on the Higgs 
mass $M_h$. Because of the
logarithmic dependence of $\lambda$ on the energy, such
considerations~\cite{con,Str,CEQ,thermal1,Hambye} can test the properties 
of the theory 
up to
extremely small distances, which are otherwise totally unaccessible to any
imaginable collider experiment.

In this paper, we want to use the same considerations for a different
purpose. Rather than trying to infer new particle-physics properties at
small distances, we will assume the validity of the Standard Model (SM) up
to an arbitrary high-energy scale, and find what information on the early
stages of the Universe can be extracted from a measurement of the Higgs
mass. We will first obtain that in the Higgs mass range $114 \gev \simlt 
M_h\simlt 130 \gev$, where the
electroweak vacuum is potentially metastable, the absence of excessive
thermal Higgs field fluctuations in the early Universe imposes a bound on the
reheat temperature after inflation $T_{RH}$. 

Then we will discuss the interplay
between the quantum Higgs fluctuations generated during the
primordial stage of inflation and the 
cosmological perturbations which  
are either currently observed   in the form of  Cosmic Microwave
Background (CMB) anisotropies,  or might be detected in the near future
in the form of tensor (gravity waves) perturbations. 
The key ingredient is that all these fluctuations depend upon the
Hubble rate during inflation and therefore, under certain assumptions, it is possible to relate the amount of Higgs fluctuations to observable properties of the CMB.
However,
excessive fluctuations of the Higgs field during inflation can pose
a threat to the stability of the present vacuum, if 
the Higgs mass lies in the metastability window.  

We will assume that the various initial inflationary patches 
of the Universe are characterized
by different microphysical parameters. In this sense
we take the point of view that the underlying theory has many vacua, 
realized in 
different patches of the Universe. This picture, 
usually referred to as the landscape \cite{landscape}, 
has been put forward especially in the context of string theory. Under this assumption, from CMB observations and from a measurement of the Higgs mass, we can derive probabilistic conclusions on the properties of our Universe.

In particular, within the  class of large-field models of inflation, we will 
compute the probability to have a Universe at the end of inflation
which both survived the quantum Higgs fluctuations and has the
right amount of observed cosmological perturbations. Such probability  
is extremely (exponentially) 
small, if the Higgs mass is below 124~GeV (for the present central value of the top mass). 
Moreover, we find that the discovery of a light Higgs together with the
detection of primordial tensor perturbations through the
$B$-mode of CMB polarization (at a level quantitatively described in sect.~6)  would imply  that we live in a very 
atypical Universe, whose  probability decreases exponentially
when the Higgs mass decreases. Such discovery could be interpreted as an indirect evidence for the existence of new physics beyond the Standard Model at some intermediate energy scale.

The paper is organized as follows. In sect.~2 we briefly review
the Higgs mass instability window. In sect.~3 the bounds on the
reheating temperature after inflation are discussed. In sect.~4
we compute the survival probability of the electroweak vacuum
during inflation and in sects.~5 and~6 we relate it to the curvature and tensor perturbations, respectively. Section~7 states our conclusions. The 
appendix contains technical details relevant to the calculation of the 
Higgs mass instability window.

\section{The Higgs mass instability window}
Let us start by reviewing the range of Higgs masses for which the
electroweak vacuum is metastable. Since our considerations refer to large
field values, for our purposes it is perfectly adequate to neglect the
bilinear term and to approximate the potential of the real Higgs field $h$
as
\beq
V(h)=\frac{\lambda (h)}{4}h^4.
\label{V}
\eeq
We will work in next-to-leading order approximation and include the
field-dependence of the quartic coupling $\lambda$, determined by the
two-loop renormalization-group (RG) equations, as well as the one-loop
corrections to the effective potential (as in ref.~\cite{CEQ}). The 
couplings $\lambda$ and the
top-quark Yukawa coupling $h_t$, entering the RG evolution, are then
related to the physical Higgs and top pole masses. The explicit
expressions are given in the appendix.

\begin{figure}
\center{
\includegraphics[width=12cm,height=9cm]{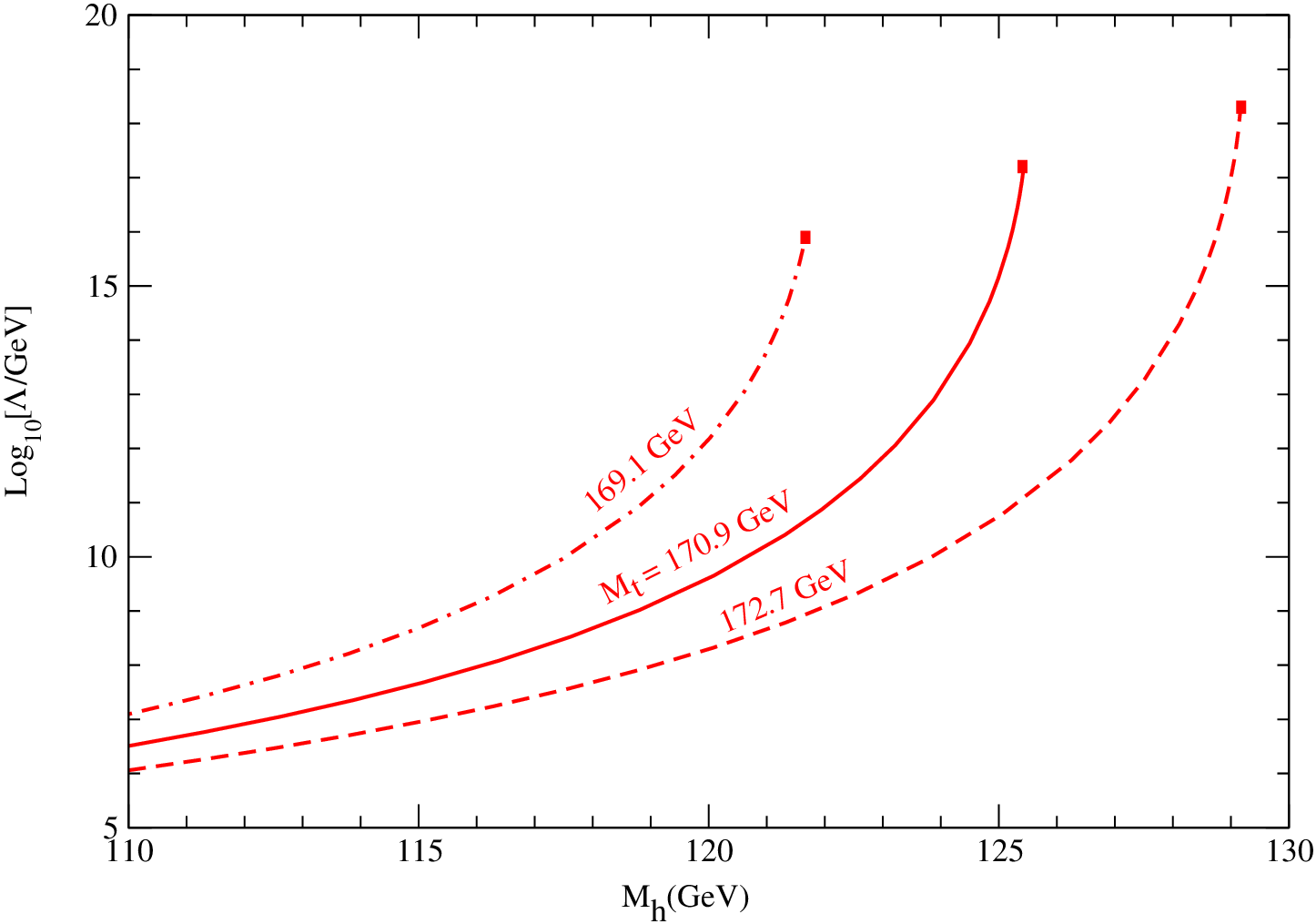}}
\caption{\label{inst} The instability scale $\Lambda$ as a function of the
Higgs mass $M_h$ for three different values of the top mass $M_t$.} 
\label{LAMBDA}
\end{figure}

The request that the electroweak vacuum $\vev{h}=v$, with 
$v=(\sqrt{2}G_F)^{-1/2}=246.22$~GeV,
is the true minimum of
the potential, up to a cut-off scale $\Lambda $, implies $\lambda( \mu
)>0$ for any $\mu <\Lambda$. This condition is not satisfied at some 
energy scale whenever $M_h$ is below some critical value 
$M_h^c$ given by
\beq
M_h < M_h^c=125.4\gev + 3.8 \gev \left( \frac{M_t -170.9 \gev}{1.8\gev} \right) 
-1.6\gev  \left( \frac{\alpha_s(M_Z)-0.1176}{0.0020}\right) \pm 2\gev .
\label{stability}
\eeq
We have explicitly
shown the dependence on the two most important SM parameters, normalizing
their effects in units of one standard deviation from their experimental
central value, taking $M_t= 170.9 \gev \pm 1.8\gev$~\cite{mtop} and
$\alpha_s(M_Z)=0.1176 \pm 0.0020$~\cite{pdg}. 
Besides the uncertainties in the SM parameters, \eq{stability} has an
overall error due to higher-order corrections which, parametrically, are
expected to shift the result by an amount $O(\alpha_s^2 M_h/\pi^2)$.
However, since the two-loop QCD correction to the top-quark pole mass
(included in our calculation and shown in the appendix) has a large
coefficient, we conservatively estimate the theoretical error to be 2~GeV.
Figure~\ref{inst} shows the instability scale $\Lambda$, at which the quartic coupling $\lambda$ becomes negative, as a function of the
Higgs mass for three different values of the top mass.  
At scales larger than $\Lambda$, the Higgs potential becomes negative, and then it develops a new minimum.
The 
endpoints of the lines, marked with a square, correspond to $M_h^c$ and to 
the scale at which the  new minimum of the Higgs potential characterized 
by a large vacuum expectation value becomes degenerate with the 
electroweak one. 

\begin{figure}
\label{window}
\center{
\includegraphics[width=12cm,height=9cm]{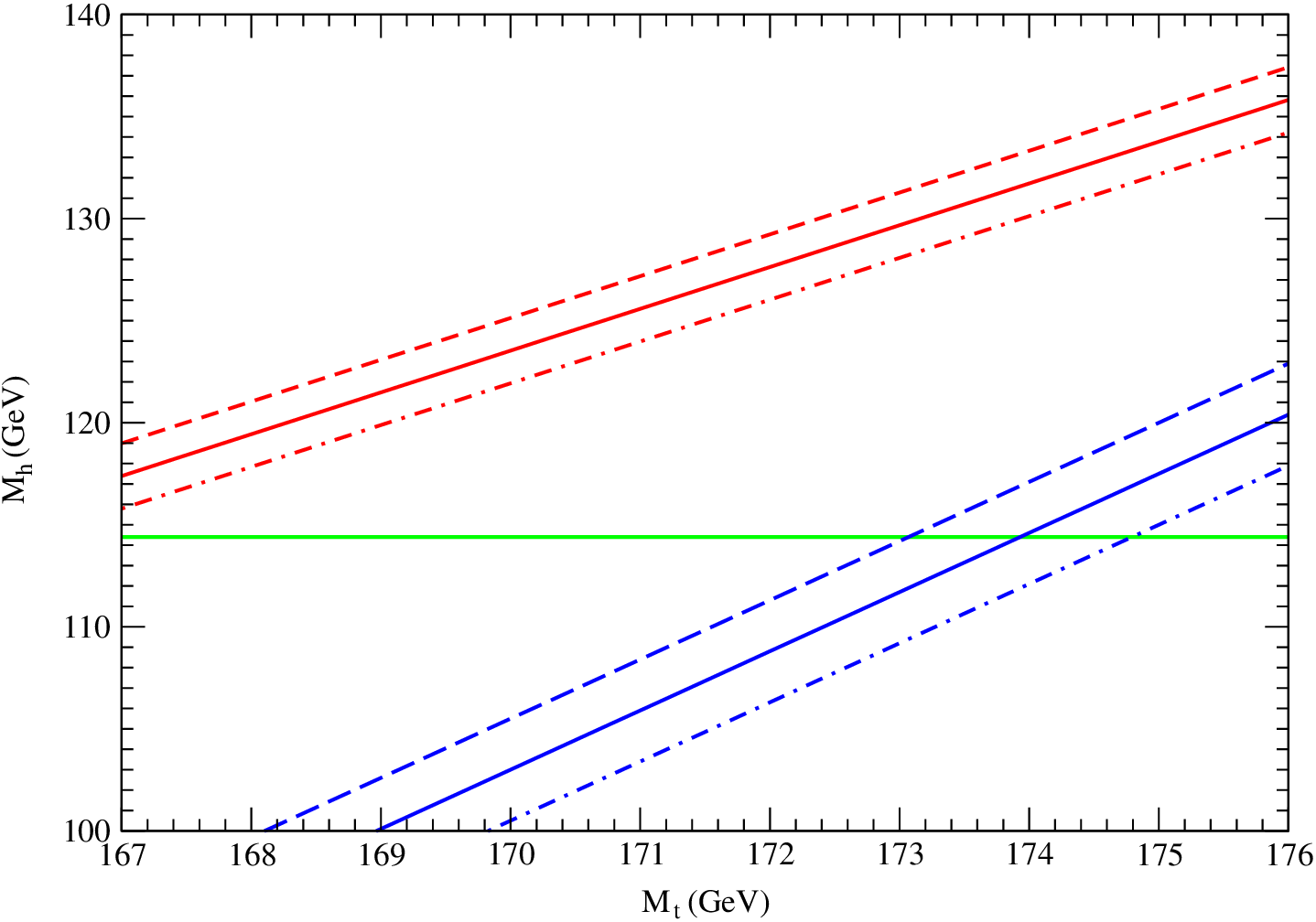}}
\caption{\label{fig:window} Lower bounds on $M_h$ from absolute stability 
(upper curves) and $T=0$ metastability (lower curves). The width 
corresponds to $\alpha_s(M_Z)=0.1176\pm 0.0020$ (with the higher curve 
corresponding to lower $\alpha_s$) and we do not show the uncertainty 
from higher-order effects, which we estimate to be below 2--3~GeV. The 
horizontal line is the LEP mass bound.} 
\end{figure}

A lower bound on $M_h$ is obtained by considering the instability of the
electroweak vacuum under quantum tunneling. Since the vacuum transition is
dominated by late times, this bound is independent of the early history of
the Universe.  The tunneling probability $p$ is given by
\beq
p=\max_{h<\Lambda} ~V_U h^4 \exp \left( 
-\frac{8\pi^2}{3|\lambda (h)|} \right)\ .
\label{transt}
\eeq
We have also included one-loop corrections to the bounce
configuration~\cite{Str}, which however have only a small impact on the
final bound on $M_h$, contributing to less than a GeV. In
\eq{transt}, $V_U$ is the space-time volume of the past light cone of the
observable Universe, and we take $V_U=\tau_U^4$, where $\tau_U$ is the
lifetime of the Universe. Taking $\tau_U=13.7\pm 0.2$~Gyrs from WMAP
data~\cite{Spergel:2006hy}, the metastability limit
$p<1$ imposes the bound on the Higgs mass \cite{Str}
\beq
M_h > 105.6\gev +5.2\gev \left( \frac{M_t -170.9 \gev}{1.8\gev} \right) 
-2.5\gev  \left( \frac{\alpha_s(M_Z)-0.1176}{0.0020}\right) \pm 3\gev .
\label{quantum}
\eeq
The error of 3~GeV is estimated by combining uncertainties from
higher-order corrections and from the prefactor in $p$.

The bound in \eq{quantum} is useful only when it is stronger than the
direct experimental limit on the Higgs mass~\cite{Barate:2003sz}
\beq
M_h > 114.4\gev ~~~{\rm at ~95\% ~CL}.
\label{lep}
\eeq
In summary, eqs.~(\ref{stability}), (\ref{quantum}) and (\ref{lep}) define
the Higgs-mass window (see fig.~\ref{fig:window}) in which the electroweak
vacuum is metastable and therefore potentially sensitive to large field
fluctuations during the early stages of the Universe. 

\section{Bound on the reheating temperature}

At sufficiently large temperatures, thermal fluctuations in the early
Universe plasma can trigger the decay of the metastable electroweak vacuum
\cite{thermal,thermal1,thermal2} by nucleation of bubbles that probe the
Higgs instability region. On the other hand, high-temperature effects also
modify the Higgs effective potential, with a tendency of making the origin
more stable. The contribution of the different plasma species to the
potential (or rather, free energy) in the non-interacting gas
approximation is given by standard one-loop (bosonic/fermionic) thermal
integrals. Each particle species, with $h$-dependent mass $M_\alpha(h)$,
contributes to the free-energy
\bea
\delta_\alpha V(h)&=&\frac{T^4}{2\pi^2}N_\alpha\varepsilon_\alpha \int_0^\infty
dx\ x^2\log\left[1-\varepsilon_\alpha
  e^{-\sqrt{x^2+M_\alpha^2(h)/T^2}}\right]\nonumber\\
&+&\frac{T}{12\pi}\frac{1+
\varepsilon_\alpha}{2}N_\alpha \left\{M^3_\alpha(h) -
  \left[ M^2_\alpha(h)+\Pi_\alpha(h,T^2)\right]^{3/2}  \right\}\
, \label{Tpot}
\eea
where $N_\alpha$ counts the number of degrees of freedom, $\varepsilon_\alpha=+1(-1)$ for bosons (fermions) and
$\Pi_\alpha(h,T^2)$ is the thermally corrected mass of the corresponding
species (see {\it e.g.} ref.~\cite{thermalmass}).  The second line in 
\eq{Tpot} takes into account the effect of
resumming hard-thermal loops for Matsubara zero modes. For our numerical
work we used a series expansion of these integrals in terms of modified
Bessel functions \cite{AH}, avoiding high $T$ expansions.

The energy $E_c(T)$ of the smallest critical bubble 
large enough to grow (overcoming the surface tension penalty) controls the
false vacuum decay rate through a Boltzmann suppression factor 
$\exp{[-E_c(T)/T]}$. The quantity $E_c(T)$ is computed by solving for the 
$O(3)$ bounce solution \cite{thermal} using the finite $T$ potential 
described above. It is easy to show \cite{thermal1} that, parametrically, 
$E_c(T)/T\sim \pi g/|\lambda(T)|$.

The vacuum decay rate per unit volume is 
\beq
\Gamma(T) \simeq T^4 \left[\frac{E_c(T)}{2\pi T}\right]^{3/2}\exp[-E_c(T)/T]\ .
\eeq
The differential decay probability $dP/d\ln T$ is obtained by multiplying 
$\Gamma(T)$ above by the volume of the Universe at temperature $T$ and the 
time spent at that $T$. In a radiation dominated Universe one has 
\beq
\frac{dP}{d\ln T}\simeq \Gamma(T) 
\tau_U^3\frac{M_p}{T^2}\left(\frac{T_0}{T}\right)^3\ ,
\eeq
where $T_0\simeq 2.73^o$ K $\simeq 2.35\times 10^{-4}$~eV and $M_p=1.2\times 10^{19}$~GeV is the Planck mass. The previous 
result assumes $T$ is smaller than the reheating temperature after 
inflation, $T_{RH}$. For temperatures $T>T_{RH}$, the previous result gets 
modified to 
\beq
\frac{dP}{d\ln T}\simeq \Gamma(T)
\tau_U^3\frac{M_p}{T^2}\left(\frac{T_{RH}}{T}\right)^{10} 
\left(\frac{T_0}{T_{RH}}\right)^3 \ ,
\eeq
to take into account the period of expansion during inflaton dominance 
before reheating is completed. The final decay probability $P$, resulting 
from integrating in $d\ln T$ the previous expressions, is not a properly 
normalized probability. Its interpretation is that the fraction of space 
converted to the true vacuum goes like $e^{-P}$.

\begin{figure}
\center{
\includegraphics[width=12cm,height=9cm]{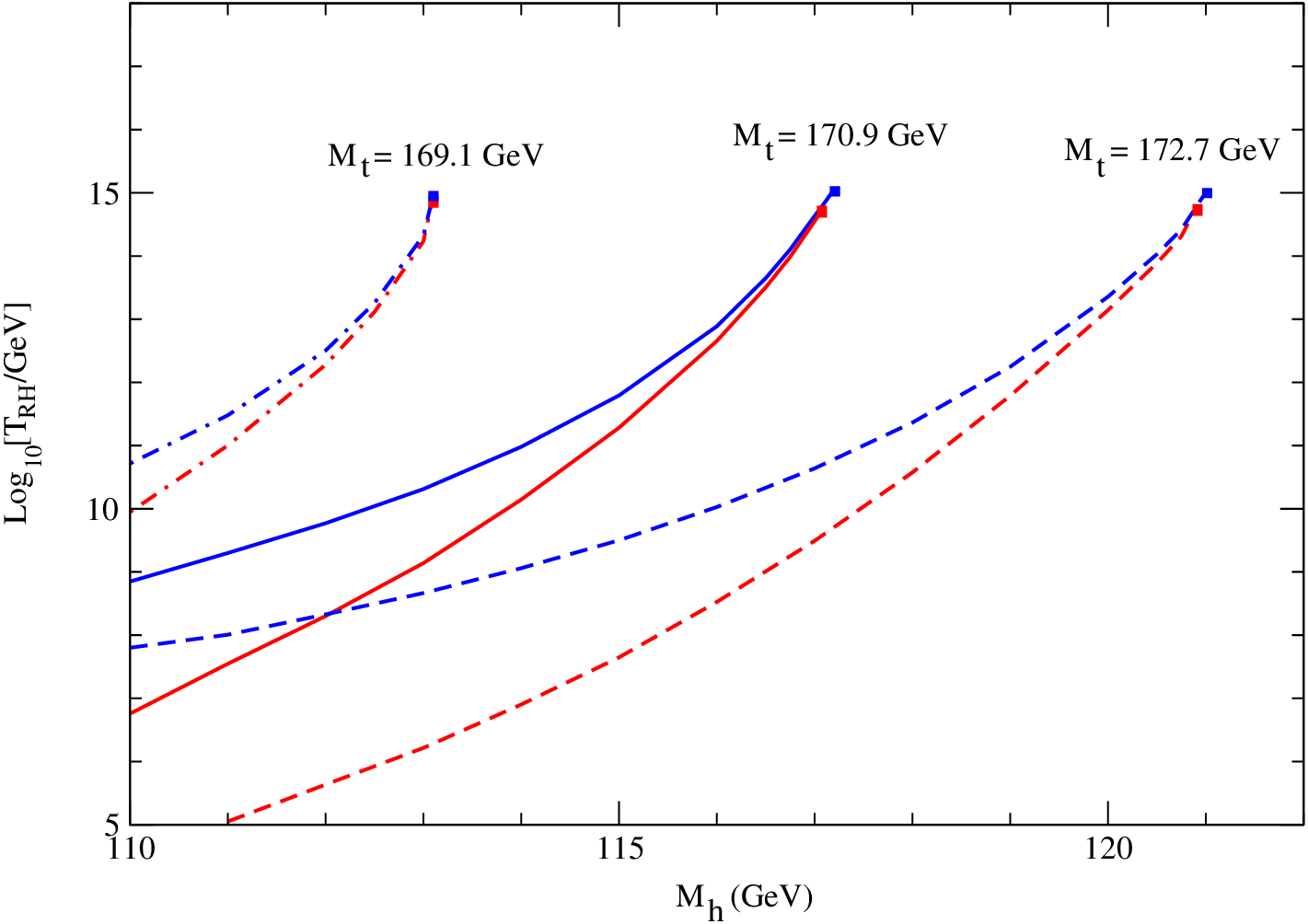}}
\caption{\label{fig:RHI} Upper bounds on $T_{RH}$, as functions of $M_h$, from sufficient 
stability of the electroweak vacuum against thermal fluctuations in the 
hot early Universe for three different values of the top mass.
The lower curves are for $H_f=10^{13}$ GeV, the upper ones for $H_f$
deduced from \eq{limtrh}, $H_f=[4\pi^3 g_* (T_{RH})/45]^{1/2}
(T_{RH}^2/M_p)$, which corresponds to the case of instant reheating.
We take $\alpha_S(M_Z)=0.1176$. Lowering (increasing) 
$\alpha_S(M_Z)$ by one standard deviation lowers (increases) the bound on 
$T_{RH}$ by up to one order of magnitude.} 
\label{mhTRH}
\end{figure}

For a given value of the Higgs mass in the metastability window of 
fig.~\ref{fig:window} the requirement that the false vacuum does not decay 
during the high $T$ stages of the early Universe (that is, $e^{-P}\simlt 
1$) will set an upper bound on $T_{RH}$. In fact $T_{RH}$ is not the 
maximal temperature achieved after inflation. Such maximal temperature 
occurs after inflation ends and before reheating completes and is given by
\cite{GKR}
\beq
\label{tmax}
T_{max}=\left(\frac{3}{8}\right)^{2/5}
\left(\frac{5}{\pi^3}\right)^{1/8}
\frac{g^{1/8}_*(T_{RH})}{g^{1/4}_*(T_{max})}M^{1/4}_{p}H_f^{1/4}T_{RH}^{1/2}
\ ,
\eeq
where $g_*(T)$ counts the effective number of degrees of freedom (with a 
$7/8$ prefactor for fermions) with masses $\ll T$ and $H_f$ is 
the Hubble parameter at the end of 
inflation. The 
metastability bound on $T_{RH}$ therefore depends on the particular value 
of $H_f$: for a given $T_{RH}$, the value of $T_{max}$ grows with $H_f$. Therefore the metastability constraint on $T_{RH}$ will be more 
stringent for larger values of  $H_f$.

Figure~\ref{fig:RHI} shows this 
metastability bound on $T_{RH}$ as a function of the 
 Higgs mass for various values of the top mass and for two choices of the
Hubble rate $H_f$ at the end of inflation. 
The lower curves correspond to $H_f=10^{13}$  GeV while the upper ones 
have $H_f=[4\pi^3 g_* (T_{RH})/45]^{1/2}
(T_{RH}^2/M_p)$, the lowest value allowed once it is required that 
 the inflaton energy density 
$\rho_\phi =3M_p^2 H_f^2/(8\pi)$ is larger than the energy density of a 
thermal bath with temperature $T_{RH}$
\beq
T_{RH} < \left[ \frac{45}{4\pi^3 g_* (T_{RH})}\right]^{1/4} M_p^{1/2} 
H_f^{1/2}.
\label{limtrh}
\eeq
This condition also corresponds to the requirement that the inflaton
lifetime $\Gamma_\phi^{-1}$ is larger than the Hubble time $H_f^{-1}$.
The bound on $T_{RH}$ gets weaker for smaller
values of the top mass (or larger values of the 
Higgs mass) since the instability scale becomes higher and 
eventually, for \cite{thermal2}
\beq
M_h > 117.4\gev + 4.2 \gev \left( \frac{M_t -170.9 \gev}{1.8\gev} \right) 
-1.6\gev  \left( \frac{\alpha_s(M_Z)-0.1176}{0.0020}\right) \pm 3\gev ,
\label{metastability}
\eeq
the bound on $T_{RH}$ is lost: the vacuum is sufficiently long-lived even 
for $T\sim M_p$ [that is, $\int^{M_p}_{T_0}(dP/d\ln T) d\ln T <1$]. 
This is the reason why the lines in fig.~\ref{fig:RHI} stop at some value 
of $M_h$. From this figure it is also clear that the bound on $T_{RH}$ is 
very sensitive on the value of $M_t$, with the experimental error in $M_t$ 
being the main source of uncertainty.

It is natural to ask what implications the bound on $T_{RH}$ shown in 
fig.~\ref{fig:RHI} has for
leptogenesis. SM thermal leptogenesis, with hierarchical right-handed
neutrinos, sets a lower bound on $T_{RH}$ as a function of $M_1$, the mass
of the lightest right-handed neutrino~\cite{thlepto}. This bound reaches
its minimum for $M_1 \sim T_{RH}$, when $T_{RH}>3\times
10^9\gev$~\cite{DI}. This condition could be in conflict with the upper
bound on $T_{RH}$ shown in fig.~\ref{fig:RHI}, if the Higgs mass turns out
to be very close to the LEP lower limit and if the top mass is on the high
side of the allowed experimental range. However we stress that these
considerations apply only to the case of hierarchical thermal leptogenesis
in the SM, with no new physics present below the scale $M_1$.

The Yukawa couplings $h_\nu$ of the heavy
right-handed neutrinos could in principle affect the bound on $T_{RH}$, since $h_\nu$ can modify
the instability scale of the
Higgs potential~\cite{CdCIQ} with its effect on the evolution of $\lambda$
above the $M_1$ threshold. Because $h_\nu^2=m_\nu M_1/v^2$, such effects
turn out to be important only if the mass of the right-handed neutrinos is
sufficiently large, $M_1\simgt (10^{13}-10^{14})\gev$ ~\cite{CdCIQ}.
Therefore, the existence of heavy right-handed neutrinos can modify the 
bounds on $T_{RH}$ we have obtained only at such large energy scales, {\it 
i.e.} for $T_{RH}>M_1\simgt \left(10^{13}-10^{14}\right)\gev$.

\section{Survival probability of the electroweak vacuum during inflation}
\noindent
In the previous section we have discussed the stability of the electroweak 
vacuum against thermal fluctuations. These are expected 
to drive the Higgs field towards the instability region 
if $T_{RH}$
is larger than the critical temperature shown in fig.~\ref{mhTRH}.
If reheating is an 
instantaneous process, so that $H_f\simeq g_*^{1/2}(T_{RH})(T_{RH}^2/M_p)$, 
then the upper bound on $T_{RH}$ (which in this case coincides with 
$T_{max}$)
can be translated into an upper bound on $H_f$. We obtain that (for 
$M_t=170.9$~GeV) 
$H_f/\Lambda \simlt 0.1$ for $M_h=115$~GeV 
and $H_f/\Lambda \simlt 10^4$  for $M_h=117$~GeV. Not only this limit on 
$H_f$ becomes 
quickly weaker as $M_h$ is
increased inside the  metastability window, but it also completely 
evaporates if reheating is a prolonged
process with $T_{max}\gg T_{RH}$. Moreover, during inflation
the Hubble rate could  be (much) larger than its value at the end of 
inflation. Therefore, even if thermal
fluctuations do not destabilize the electroweak vacuum,  the ratio 
$H/\Lambda$ during inflation can be much
larger than one.
It is thus interesting to
address the issue of what is the fate of the electroweak vacuum
when we account for the Higgs fluctuations generated during inflation,
whose amplitude
is directly proportional  to the Hubble rate and can probe the instability 
region.

In the inflationary cosmology picture, the present structure in the 
Universe is supposed to originate from vacuum fluctuations which were 
quantum-mechanically excited during an inflationary stage 
\cite{guth,lrreview}. Indeed, any scalar field whose mass is lighter than 
$H$, where $H$ denotes the Hubble rate during inflation, gives rise to an 
almost scale-invariant spectrum of perturbations on superhorizon scales 
\cite{lindebook,tonireview}. The Higgs field, if light enough during 
inflation, does not represent an exception. We will start by considering 
the case in which the Higgs field is minimally coupled to gravity and 
therefore it can be considered nearly massless during inflation, as long 
as $H \gg M_h$. At the end of sect.~\ref{sec4} we will discuss the interesting case in which there is 
a direct coupling between the Higgs bilinear and 
the Ricci scalar.

If the Higgs mass lies in the metastable window (114~GeV $<M_h \simlt$ 
130~GeV), excessive fluctuations of the Higgs field may pose a threat to 
the stability of the electroweak vacuum.  In the inflationary picture, 
long wavelength perturbations of the Higgs field may be generated and they 
behave in the same way as a homogeneous classical field. We suppose that 
in the beginning of inflation the whole inflationary domains are 
characterized by a very homogeneous Higgs field with vanishing vacuum 
expectation value. Then the domains become exponentially large. The Higgs
quantum fluctuations divide the Universe into 
exponentially large regions with different values of the Higgs field: the 
Universe becomes quickly filled with domains in which $h$ 
typically changes from $-\langle h^2\rangle^{1/2}$ to $\langle 
h^2\rangle^{1/2}$. Here $\langle h^2\rangle$ is the variance of the Higgs 
field which we will estimate in the following.

The process of generating a classical Higgs field configuration
in the inflationary Universe can be interpreted as the
result of the Brownian motion of the Higgs scalar field under the
action of its quantum fluctuations which are converted into
the classical field when their wavelengths overcome the horizon
length. The best way to describe the structure of the Higgs fluctuations 
is provided by the stochastic approach in which one defines
the distribution of probability $P_c(h,t)$ to find the Higgs field value 
$h$ at a given time at a given point \cite{star}. Perhaps a more 
adequate approach
might be based on the distribution of probability
$P_p(h,t)$ to find the Higgs field value $h$ at a given time in a given 
physical
volume \cite{phys} which takes into account the exponential
growth of the volume of the various domains. Both distributions of 
probabilities are plagued by various problems (for
a recent discussion on this issue  related to the idea of eternal 
inflation, see ref.~\cite{linep}). The comoving probability $P_c$ at a given point
ignores the possible creation of new volumes (new points) during 
inflation; the physical probability suffers from normalizability  
problems. 
In this paper we infer our results  by making use
of the comoving probabiliy $P_c$ which is technically more feasible.
The subscript $c$ serves to indicate that $P_c$ corresponds to the 
fraction of original {\it comoving} volume filled by the Higgs field
$h$ at the time $t$. The comoving probability 
satisfies the Fokker-Planck equation
\begin{equation}
\label{fp}
\frac{\partial P_c}{\partial t}=\frac{\partial}{\partial h}\left[
\frac{H^3}{8\pi^2}\frac{\partial P_c}{\partial h} 
+\frac{V'(h)}{3H}P_c\right]\, .
\end{equation}
In writing this  equation we have supposed that the Higgs field gives
a negligible contribution to the energy density of inflation and that the
probability does not sensitively depend upon the value of the inflaton 
field. The first assumption always holds in the relevant region of Higgs field configurations and  we will come back to the second point later on.

To solve \eq{fp} we assume that 
the Hubble rate is approximately constant during inflation. We denote from 
now on by 
$\Lambda$ the value
of the Higgs field at which the maximum of the potential $V(h)$ occurs. It
is very close to the value of the energy at which
the Higgs potential vanishes
shown in 
fig.~\ref{LAMBDA}. In those 
patches where the Higgs field takes any value larger than 
$\Lambda$, the Higgs field rapidly rolls  down towards 
the region where the potential
is negative, $V(h)= -\left|\lambda(h)\right|h^4/4$. Therefore, the domain
where the barrier has been surmounted 
will correspond to bubbles of AdS space which will rapidly collapse and disappear. 
To account for this process we impose that the solution of  
\eq{fp} satisfies the boundary condition
\begin{equation}
\label{bc}
P_c(\Lambda,t)=0.
\end{equation}
Physically, this condition dictates that, once the unstable region
of the Higgs potential is probed, the probability to jump back to the stable
region vanishes. As $\partial P_c/\partial h$ does not vanish at $h=\Lambda$,
the probability that the Higgs field is in the stable region $h<\Lambda$
will decrease with time. Since $P_c(h,t)$ gives the fraction of the comoving 
volume occupied by regions inside which the Higgs field takes the value
$h$ at a given time $t$, we expect that this fraction will become smaller and
smaller as time goes by during the inflationary stage.

To investigate the amount of Higgs fluctuations during inflation, we 
can also parametrize them 
in terms of the Higgs correlations determined with the help of the
comoving probability given by \eq{fp}.
\begin{equation}
\label{chain}
\frac{d}{dt}\langle h^m\rangle =  \frac{H^3}{8\pi^2}
m(m-1)\langle h^{m-2}\rangle-\frac{m}{3H}\langle h^{m-1}V'(h)\rangle    
+\frac{H^3\Lambda^m}{4\pi^2}P^\prime_c (\Lambda)
\, ,
\end{equation}
where $m$ is an even integer, the  prime denotes derivatives with respect 
to $h$, and
\begin{equation}
\langle h^m\rangle=\int_{-\Lambda}^{\Lambda}\, dh~h^m P_c(h,t)\ . 
\end{equation}
We consider the case in which $P_c (h,t)$ is an even function of $h$, so 
that any Higgs correlation functions with odd $m$ identically vanishes.

For constant $H$, \eq{fp} can be solved by separation of variables
\beq
P_c(h,t)=\sum_{n=0}^\infty c_n e^{-\left( \alpha 
V+\frac{H^3a_nt}{8\pi^2}\right)}\Phi_n(h) ,
~~~~~\alpha=\frac{8\pi^2 }{3H^4}\ ,
\label{solexp}
\eeq
where $\Phi_n$ and $a_n$ are the eigenfunctions and the eigenvalues of 
the equation
\beq
\Phi_n^{\prime\prime}-\alpha V^\prime \Phi_n^{\prime}=-a_n \Phi_n\ .
\label{eigen}
\eeq
Since $V(h)$ is an even function of $h$ and \eq{eigen} does not 
mix even and odd eigenfunctions, we look for solutions which are even, 
in the range $-\Lambda <h<\Lambda$. Moreover, since the condition in 
\eq{bc} has to be satisfied at any time $t$, we have to impose $\Phi_n 
(\Lambda) =0$, for any $n$. 

We assume that the Higgs field
is initially localized at $h=0$, $P_c(h,0)=\delta(h)$, and study its evolution. 
We have solved the 
Fokker-Planck equation numerically, but it is useful to give here 
approximate analytic solutions in order to describe our results.

{\it (i) Case} $H \gg \Lambda$.
Let us first consider the most interesting case in which the Hubble rate 
is much larger than $\Lambda$. In this case the potential term in 
\eq{eigen} can be neglected since $\alpha V^\prime 
\Phi^\prime_n/\Phi_n^{\prime\prime} \simlt \lambda (\Lambda /H)^4 \ll 1$, 
once we use the condition $|h|<\Lambda$. This means that the loss of 
probability caused by the AdS instability dominates the dynamics, while 
the effects from the Higgs potential are negligible. Then the solution of 
the Fokker-Planck equation satisfying the appropriate boundary conditions 
is\footnote{One can show that $P_c(h,0)=\delta(h)$ using 
$\sum_{n=0}^\infty \cos [(n+1/2)x] =\pi \delta (x)$.} 
\beq
P_c(h,t)=\frac{1}{\Lambda}\sum_{n=0}^\infty e^{-\left( n+\frac 
12\right)^2 \frac{H^3t}{8\Lambda^2}}\cos \left[ \left(n+\frac 12 \right) 
\frac{\pi h}{\Lambda}\right] .
\label{sollib}
\eeq
Notice that, in the limit of large $\Lambda$, the probability 
distribution in \eq{sollib} reduces to a Gaussian
\beq
\lim_{\Lambda \to \infty}
 P_c(h,t)=\sqrt{\frac{2\pi}{H^3t}}e^{-\frac{2\pi^2h^2}{H^3t}},
\eeq
as it can be easily obtained by switching from discrete to continuous 
variables $(n+1/2)\pi /\Lambda \to k$ and by integrating over $k$.

The survival probability $P_\Lambda$ for the Higgs to remain in the 
region $|h|<\Lambda$ and the variance of the Higgs field are given by
\beq
P_\Lambda(t)\equiv \int_{-\Lambda}^{\Lambda}\, dh~ P_c(h,t) =
\frac{2}{\pi}\sum_{n=0}^\infty \frac{(-)^n}{n+\frac 12}
e^{-\left( n+\frac 12\right)^2 \frac{H^3t}{8\Lambda^2}} ,
\label{sol1}
\eeq
\beq
\vev{h^2}\equiv \int_{-\Lambda}^{\Lambda}\, dh~ h^2P_c(h,t) =
\frac{2\Lambda^2}{\pi}\sum_{n=0}^\infty \frac{(-)^n}{n+\frac 12}\left[ 
1-\frac{2}{\left(n+\frac 12\right)^2 \pi^2}\right]
e^{-\left( n+\frac 12\right)^2 \frac{H^3t}{8\Lambda^2}} .
\label{sol2}
\eeq

At small $t$, eqs.~(\ref{sol1}) and (\ref{sol2}) become
\bea
P_\Lambda(t) \simeq 1-\sqrt{\frac{H^3t}{2\pi^3\Lambda^2}} 
e^{-\frac{2\pi^2\Lambda^2}{H^3t}} ~~~~~~&{\rm (small}~ t{\rm 
)}& ,\label{sol3}\\
\vev{h^2}\simeq \frac{H^3t}{4\pi^2}~~~~~~~~~~~~~~~~~~~&{\rm (small}~ 
t{\rm )}& .\label{sol4}
\eea
At the initial stages, the loss of probability is very small and 
$P_\Lambda$ is exponentially close to one, while $\vev{h^2}$ starts 
growing linearly with time. To study how fast the probability decays with 
time, we can consider eqs.~(\ref{sol1}) and (\ref{sol2}) in the limit of 
large $t$,
\bea
P_\Lambda(t) \simeq \frac{4}{\pi} e^{-\frac{H^3t}{32\Lambda^2}} 
~~~~~~~~~~~~~~~~~~~~~~~~~~~~~&{\rm (}H\gg \Lambda{\rm ,~large}~ t{\rm 
)}& ,\label{sol5}\\
\vev{h^2}\simeq \frac{4(\pi^2-8)\Lambda^2}{\pi^3}e^{-\frac{H^3t}{32\Lambda^2}} 
~~~~~~~~~~~~~~~~~~~&{\rm (}H\gg \Lambda{\rm ,~large}~ t{\rm )}&.\label{sol6}
\eea
The survival probability $P_\Lambda$ and the Higgs correlation function 
exponentially decay with time. The normalized Higgs variance, which 
measures the correlation function within the surviving domains, tends 
asymptotically to a constant,
$\vev{h^2}/P_\Lambda \to (1-8/\pi^2)\Lambda^2\simeq (0.44~\Lambda)^2$. The 
asymptotic behavior shown 
in eqs.~(\ref{sol1}) and (\ref{sol2}) is valid for $t \gg \Lambda^2/H^3$ 
and therefore it is reached very rapidly, justifying the assumption of 
neglecting the time dependence of the Hubble constant.  

From \eq{sol2} we can also obtain the maximum value of the Higgs variance
\beq
\vev{h^2}_{\rm max} \simeq\left( 0.39~\Lambda \right)^2 ,
\eeq
which is obtained at the time $t\simeq 10.6~\Lambda^2/H^3$.

{\it (ii) Case} $\Lambda \gg H$.
At the first stages of inflation, the first term in \eq{eigen} dominates and the results given in eqs.~(\ref{sol3}) and (\ref{sol4}) for small $t$ are valid also in the case $\Lambda \gg H$. 
At later times the shape of the probability distribution changes because the second term
in \eq{eigen} becomes relevant.
To estimate when the effect of the Higgs potential is important, we 
study \eq{chain} for the variance of the Higgs field ($m=2$)
\beq
\frac{d}{dt}\vev{h^2} =\frac{H^3}{4\pi^2} \left[ P_\Lambda +\Lambda^2 P_c^\prime (\Lambda )\right] -\frac{2\lambda}{H}{\vev{h^2}}^2 .
\label{chain2}
\eeq
Here we have neglected the contribution to the
correlators from the Higgs dependence in the Higgs quartic coupling, and we
have adopted a Hartree-Fock approximation for the nonlinear term by taking
$\vev{hV^\prime (h)}\simeq 3 \lambda {\vev{h^2}}^2$, where $\lambda$ is evaluated at the scale $\sqrt{\vev{h^2}}$.
From \eq{chain2}  we deduce that the Higgs variance starts
growing linearly with time, $\langle h^2\rangle = H^3 t/(4\pi^2)$, and
then it reaches a maximum value which can be
estimated by equating the right hand side of \eq{chain2} to zero
\begin{equation}
\label{varmax}
\vev{h^2}_{\rm max} \simeq \frac{H^2}{2\pi\sqrt{2\lambda}}\ . 
\end{equation}
This value   is reached at a number of e-foldings 
approximately given by $N\sim Ht\sim  \pi \sqrt{2/\lambda}$. Therefore
this maximum is reached very promptly which, also in this case, justifies 
neglecting the
time dependence of the Hubble rate. 
At times $t\simgt 1/(\sqrt{\lambda}H)$, the friction term starts
being relevant in \eq{fp}.  
 In the absence
of the AdS instability region, the comoving probability  would reach a 
stationary form  $P_c(h)\propto \exp [-8\pi^2 V(h)/3H^4]$, {\it i.e.} 
$a_0=0$ and $\Phi_0=1$ in \eq{solexp}.
However, in the presence of the AdS region, this solution has to be 
modified to account for the loss of the probability. The stationary 
solution does not satisfy the boundary condition
in \eq{bc} but, for $\Lambda \gg H$, $P_c(\Lambda)$ is exponentially 
close to zero. Therefore we can obtain
the correct solution by perturbing around the stationary solution valid 
for $\Lambda \to \infty$, and we find
$\Phi_0(h) \simeq 1-I(h)/I(\Lambda)$ and $a_0\simeq 1/I(\Lambda)$, where
\beq
I(h)=\int_0^h dh^\prime e^{-\alpha V(h^\prime)}  \int_{h^\prime}^h 
dh^{\prime \prime} e^{\alpha V(h^{\prime \prime})}\ .
\label{defi}
\eeq
At late times, the survival probability decays with time as 
$P_{\Lambda}(t)\propto e^{-\gamma t}$, with
\beq
\gamma = 
\frac{H^3a_0}{8\pi^2}\simeq \frac{2\sqrt{\pi}}{\Gamma (1/4)}
\left( \frac{2\lambda}{3}\right)^{5/4} 
\frac{\Lambda^3}{H^2}e^{-\frac{8\pi^2V(\Lambda)}{3H^4}}
~~~~~{\rm (}\Lambda\gg H{\rm ,~large}~ t{\rm )}\ .
\label{asyp}
\eeq
Here we have integrated \eq{defi} neglecting the field 
dependence of the quartic Higgs coupling $\lambda$ and we have taken the limit $\Lambda\gg H$.
In the more relevant case in which the Hubble rate is changing with time, we
have numerically checked that the survival probability decays as
$\exp [-\int^t dt' \gamma(t')]$, where one can approximately use
the expression of $\gamma$ in \eq{asyp} simply plugging the
appropriate time-dependence of the Hubble rate. This is also true in the case $H\gg \Lambda$, using the time
dependence shown in \eq{sol5}.

\section{The survival probability and the comoving curvature perturbation}
\label{sec4}

\noindent
From the results of the previous section we deduce that, 
if $H\gg \Lambda$, an exponentially small
fraction of the initial comoving volume survives the Higgs quantum fluctuations
during inflation. The number of surviving domains will be suppressed by
${\rm exp}\left(-H^2 N/32 \Lambda^2\right)$, where $N$ is the 
total number of e-folds.  
May we conclude that our Universe is 
very unlikely?  The answer is no. Suppose, indeed, that to quantify the 
probability of ending up in a ``well-behaving'' domain (the vacuum has 
the correct Higgs vacuum expectation value) we simply count the fraction 
of domains where the value of the Higgs field is less than 
$\Lambda$. This fraction is 
given by the comoving survival probability. However, those regions where 
the value of the Higgs field has become  larger than the critical 
value during the
inflationary stage simply do 
not exist. As soon as the instability point is reached, those regions 
collapse and disappear. Only the ``well-behaving'' regions remain and 
their corresponding fraction is unity.

Can we then conclude that the parameter space where $P_\Lambda$ is
exponentially small is acceptable?
As we will show, the answer is again no, at least in a 
probabilistic sense. Indeed, 
in the class
of inflationary models dubbed large-field models,
the ``well-behaving'' domains, where the vacuum expectation value of the 
Higgs is correct, typically have an insufficient amount 
of curvature perturbations generated during inflation. In the 
inflating domains that are characterized by the correct amount of 
curvature perturbations, the Hubble rate, and therefore 
the square root of the Higgs variance, turns out to 
be much larger than the scale $\Lambda$. Therefore, 
among the surviving regions, the ones having the right amount of
curvature perturbations will be exponentially rare. 

Observations of the CMB   
anisotropies are consistent with a smooth 
and nearly Gaussian power spectrum of curvature perturbations with an 
amplitude 
\beq
\zeta_{\rm obs} \simeq 5\times 10^{-5}.
\eeq
Inflation predicts such a spectrum with amplitude
\begin{equation}
\label{k}
\zeta=\frac{H_{*}^2}{2\pi \dot{\phi}_{*}},
\end{equation}
where $H_{*}$ and $\phi_{*}$  are  the value of the Hubble rate and the
inflaton field, respectively, when there are about 60 e-folds
to go till the end of inflation. The dot stands for differentation with respect
to time.

Let us focus on the large-field models of inflation. 
We parametrize their potential by 
\begin{equation}
V(\phi)=\frac{\mu^{4-p}}{p}\phi^p,
\label{potf}
\end{equation}
with $p$ a 
positive integer. This determines the amplitude of the curvature perturbations
\begin{equation}
\zeta=
\left(\frac{4N_*}{\pi p}\right)^{1/2}
\frac{H_*}{M_p},
\label{phi}
\end{equation}
where $N_*$ is  the number of e-folds till the end of inflation
when the observable scales leave the horizon\footnote{The number 
of e-folds to go till the end
of inflation for those scales which exit the horizon and are today relevant
for observation 
depends on the reheating temperature $T_{RH}$ through the
relation $N_*\simeq 60 +1/6\ln(-n_T)+1/3 \ln(T_{RH}/10^{16}\,{\rm GeV})-1/3
\ln\,\gamma$ \cite{lrreview},
 where $n_T$ is the spectral index of tensor perturbations
and $\gamma$ is the ratio of the entropy per comoving volume
today to that after reheating.}. Notice, in particular, that
the perturbation is directly proportional to the Hubble rate. This will play
a crucial role in the following because larger values of the Hubble rate, and
therefore larger values of $\zeta$, tend to destabilize the electroweak vacuum.

In the case in which $H\gg \Lambda$, the survival probability
becomes
\begin{equation}
\label{ll}
P_\Lambda\propto \exp \left[{-\int_0^t dt' \frac{H^3(t')}{32\Lambda^2}}\right] \simeq
\exp\left[ {-\frac{ H_i^2 N}{16  (p+2)\Lambda^2}} \right] ,
\end{equation}
where $H_i$ is the Hubble rate
at the beginning of inflation 
and $N$ is the total number of e-folds. 
Using \eq{phi} and $(H_i/H_*) = (N/N_*)^{p/4}$,
the survival probability can be expressed in terms of the total
number of e-folds and the comoving curvature perturbation
\begin{equation}
P_\Lambda\propto \exp \left[ {-\frac{\pi p}{p+2}
\left(\frac{\zeta M_p}{8\Lambda}\right)^2
\left( \frac{N}{N_*}\right)^{\frac{p+2}{2}}}\right] .
\label{probdist}
\end{equation}

We now take a crucial step and assume  
that initially 
the Universe is characterized by  local domains where 
the inflationary parameters may assume any value. 
As mentioned in the introduction, this assumption is inspired by the string landscape picture~\cite{landscape},
in which the theory has an enormous number of vacua, each determining different values for the underlying parameters. This hypothesis agrees especially well with the point of view followed in this paper of ignoring the hierarchy problem associated to the Higgs mass and extending the validity of the SM up to very high-energy scales. Indeed, this approach could be justified in the context of a landscape scenario in which the Higgs mass parameter scans among the different vacua.
In practice, in the case under consideration, we allow for
the possibility that the various initial comoving patches 
of the Universe are characterized
by different microphysical parameters, such as the parameter 
$\mu$ in \eq{potf} and
the initial value
of the inflaton field $\phi_i$. These two parameters
determine the other cosmological variables
\beq
N=\frac{4\pi \phi_i^2}{pM_p^2},~~~~H_i^2=\frac{8\pi \mu^{4-p}\phi_i^p}{3pM_p^2},~~~~
\zeta =\sqrt{\frac 23}\frac{N_*^{\frac{p+2}4}}{\pi^{\frac p4}}\left( \frac{2\mu}{\sqrt{p}M_p}\right)^{\frac{4-p}2}.
\label{variab}
\eeq
Using \eq{variab}, we trade the  parameters $\phi_i$ and $\mu$ for the two 
more physical quantities $N$ and $\zeta$.
The way the inflationary parameters  are distributed at the beginning of 
inflation 
among these different 
regions is unknown and we take it to be simply flat in $N$ and $\zeta$. 
Our results are not very sensitive to this assumption (as long as the 
parameter distributions
are not sharply peaked), because the  probability function we are 
interested in has an exponential dependence on our variables. 
Among the  comoving 
patches which have survived the Higgs instability, the probability
that the total number of e-folds is between $N$ and $N+dN$ and that the 
comoving curvature perturbation is between $\zeta$ and $\zeta+d\zeta$ is 
$P_\Lambda(\zeta,N)\ d\zeta\ dN$, with $P_\Lambda(\zeta,N)$ given in 
\eq{probdist}.

The range in which the parameters scan is in principle arbitrary. We can 
restrict it using anthropic priors, although this is not essential for 
our conclusions. 
If perturbations are too large,
$\zeta\simgt 10^{-4}$, they become non-linear when the average
density of the Universe is too high  to allow the resulting structures to 
guarantee a stable environment for
life \cite{anth}. On the opposite side,  if the perturbations are too 
small, $\zeta\simlt 10^{-6}$, the
majority of the overdense regions are not  capable of cooling quickly 
enough 
to form structures \cite{anth}. 
Therefore, we scan the comoving curvature perturbations
only over the anthropically allowed region $10^{-6}\simlt
\zeta\simlt 10^{-4}$. 

\begin{figure} 
\label{logP1}
\center{ \includegraphics[width=12cm,height=9cm]{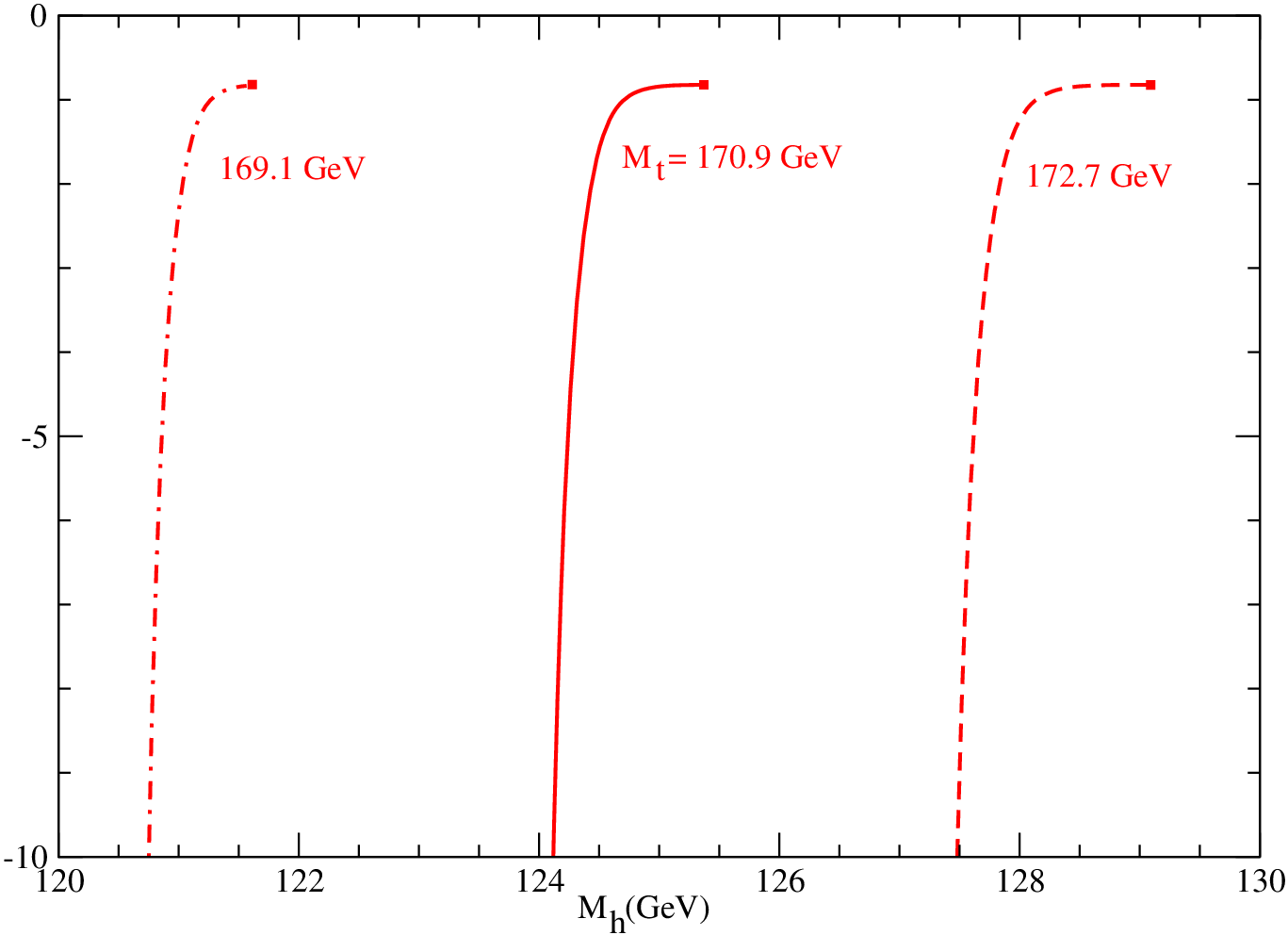}}
\caption{\label{fig:Lintprob} The quantity 
$\log_{10}\left[\int_{\zeta>\zeta_{\rm obs}}d\zeta \,P_\Lambda/
\int_{\zeta>10^{-6}}d\zeta \,P_\Lambda\right]$
as a function of the Higgs mass for three values of $M_t$ and $p=2$.} 
\end{figure}

The probability in \eq{probdist} is dominated by the smallest value of $N$, which we
choose to be $N_*$ 
in order to allow at least sixty e-folds of inflation.
Integrating over the total number of e-folds, the 
probability becomes 
\begin{equation}
P_\Lambda(\zeta)\propto E_{\frac{p}{p+2}} (\alpha )\simeq
\frac{e^{-\alpha}}{\alpha},~~~~\alpha \equiv \frac{\pi p}{p+2}
\left(\frac{\zeta M_p}{8\Lambda}\right)^2.
\end{equation}
Here  ${\rm E}_n(x)$ is the exponential 
integral function and we have expanded for large values of the
parameter $\alpha$, which is appropriate since \eq{probdist} is valid only for $H\gg \Lambda$ ({\it i.e.} $\alpha \gg 1$).
When $\alpha <1$, $ P_\Lambda(\zeta)$ should be replaced by a constant.
The probability $P_\Lambda(\zeta)$ is peaked at the smallest
value of the cosmological perturbation $\zeta$.  Therefore, the fraction of domains
whose value of the cosmological perturbation is as large as the measured
$\zeta_{\rm obs}$ is extremely small, even if we impose the prior of restricting our considerations only to values of $\zeta$ within the narrow anthropic range:
\begin{equation}
\frac{\int_{\zeta>\zeta_{\rm obs}}d\zeta \,P_\Lambda(\zeta)}{
\int_{\zeta>10^{-6}}d\zeta \,P_\Lambda(\zeta)}\simeq
\left(\frac{10^{-6}}{\zeta_{\rm obs}}\right)^3
 {\rm exp}\left[-
\frac{\pi p}{p+2}
\left(\frac{\zeta_{\rm obs} M_p}{8\Lambda}\right)^2
\right] .
\label{trent}
\end{equation}
We recall that \eq{trent} is valid for $H\gg \Lambda$ which 
holds in the anthropic window for $\zeta$ as long as $\Lambda <10^{12}$~GeV.
In fig.~(\ref{logP1}) we plot this ratio as a function of the
Higgs mass and for different values of the top mass and for $p=2$. 
As expected, if
the Hubble rate $H_*$ is large enough to account for $\zeta$,
see \eq{phi}, the fraction of volumes which
survived the quantum Higgs fluctuations is extremely small, unless the mass of the
Higgs is close to (or larger than) the critical $M_h^c$ of 
\eq{stability} shown by the upper curves in fig.~2.

Let us now discuss how our results change if we adopt a physical
distribution $P_p(h,t)$ instead of the comoving one $P_c(h,t)$.
As we already mentioned, $P_p(h,t)$  describes the probability
of finding the Higgs field value $h$ at a given time in a physical volume 
and takes into
account the exponential growth of the volume of the various inflationary 
patches. This amounts to multiplying the comoving probability by 
${\rm exp}(3 N)$, where $N$ is the total number of e-folds. Putting aside the
technical problems related to the normalizability of $P_p$, 
we can express
the physical survival probability $P_{\Lambda,p}$ as the product of 
the exponential volume factor with $P_\Lambda$ given in  \eq{ll}
\begin{equation}
P_{\Lambda,p}\propto \exp\left[ {-\frac{ H_i^2 N}{16  (p+2)\Lambda^2}}+3N 
\right].
\end{equation}
From this expression it is clear that, when $H_i^2\simgt 48(p+2) \Lambda^2$, 
the
term in the exponential accounting for the dynamics of the Higgs  
fluctuations dominates
over the one due to the physical volume expansion. 
Also, when $\zeta$ and $N$ are used as independent scanning variables, 
see \eq{probdist}, 
the Higgs term $\exp [-\alpha (N/N_*)^{1+p/2}]$ has a steeper dependence 
on $N$ than the
volume term $\exp (3N)$, when $N$ grows.
This implies
that even the  physical survival probability is pushed towards
small values of the number of e-folds $N$ and our results are not 
modified by using $P_p$ rather
than $P_c$. Because we have found that statistics prefer small $N$ (or 
equivalently small $\phi_i$), we can
avoid problems related to the 
computation of the physical probability when this is maximized for large
values of the inflaton field. Indeed, for values of the inflaton field 
$\phi \simgt \mu
(M_p/\mu)^{6/(p+2)}$, the inflaton 
evolution  is quantum rather than classical giving rise to 
eternal inflation \cite{phys}. The computation of the
physical probability to get a certain value of the cosmological perturbation
$\zeta$ in a physical volume would become therefore a complicated task as 
the Brownian motion of the inflaton field
described by the Fokker-Planck equation has to be taken into account. 
In ref.~\cite{FHWdp} it is argued that the volume factor $\exp (3N)$ 
statistically favors small values of $\zeta$. This conclusion is based on 
the requirement that the inflaton evolution remains classical, and this 
determines a maximum value of the number of e-folds $N_{max}\sim 
(M_p/\mu)^{2(4-p)/(p+2)}\sim \zeta^{-4/(p+2)}$, which selects small 
$\zeta$. This requirement is not  well motivated on physical grounds and 
the quantum regime might modify the parametric dependence of the 
probability on the perturbation $\zeta$. On the contrary, in our case the Higgs fluctuations enhance the probability of domains with small values of $\phi_i$, away from the region dominated by quantum evolution.

So far, we have discussed the case in which 
 inflation is driven 
by an inflaton field with a polynomial potential. There are two other classes
of inflationary models, called small-field and hybrid models \cite{lrreview},
where the potential is made of a constant vacuum energy 
plus a field-dependent term responsible for the slow-roll. In such a case 
the connection between the Hubble rate and the cosmological perturbation 
is lost. For instance, in a model with potential
$V(\phi)=V_0+m^2\phi^2/2$, the Hubble rate during inflation is roughly
$H_*\sim(\zeta\phi_f/N_*)$, where $\phi_f$ is the value of the
inflaton field at which inflation stops. Therefore, one can have a large
probability of having domains
with the correct amount of perturbation, but values of
the Hubble rate easily smaller than the instability point $\Lambda$ 
by simply choosing $\phi_f$ small enough. However, as will be discussed in the next section, if a future
measurement of gravity waves through the $B$-mode of the CMB polarization 
indicates that the Hubble rate is sizeable, even within 
these classes of models, one would have to conclude that we live
 in an unlikely Universe. 

\begin{figure} 
\label{conformal}
\center{ \includegraphics[width=12cm,height=9cm]{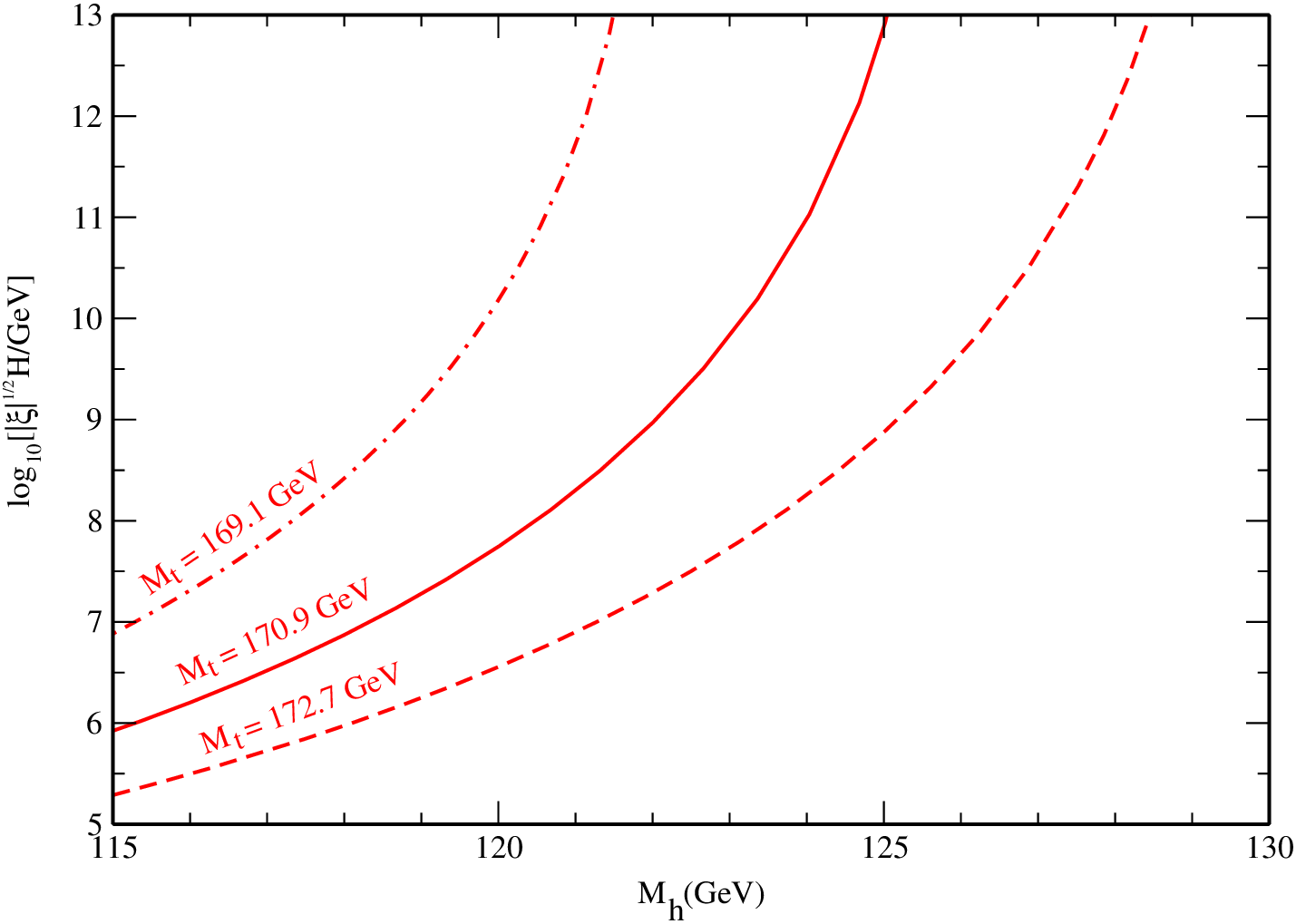}}
\caption{\label{fig:conformal} Upper bound on $\sqrt{|\xi|}H$ as a function of $M_h$ obtained by requiring
that the electroweak vacuum is stable. Here $\xi$ is the {\it negative} coupling between the Higgs bilinear and the scalar curvature, and $H$ is the maximal Hubble rate during inflation.} 
\end{figure}

Finally, we study the case in which we relax  
the assumption that the Higgs is
minimally coupled to gravity.  Indeed, the Higgs field might be coupled
nonminimally to gravity through a term in the Lagrangian $\xi R h^2/2$,
where $\xi$ is a numerical coefficient and $R$ is the Ricci scalar. During
inflation this term gives a contribution to the Higgs mass equal to $12\xi
H^2$ (since $R=-12H^2$). If $0<\xi\ll 10^{-1}$, the Higgs field is
practically massless during inflation and our conclusions do not change.
However, when $|\xi|$ is sizable, our results are drastically modified.  
If $\xi \gsim 10^{-1}$, the electroweak vacuum
is stable and the Higgs fluctuations are damped exponentially because of
the large positive effective mass. However,
if $\xi<0$ (as in the case of conformal coupling, where $\xi=-1/6$), the
electroweak vacuum is destabilized by a large tachyonic mass even without
inflationary fluctuations.  In fig.~\ref{conformal}
we show the upper bound on $\sqrt{|\xi|}H$, in the case of negative $\xi$, obtained by imposing that
the electroweak vacuum is not destabilized by the large
negative mass squared induced by the  $\xi R h^2/2$ term. Here $H$ has to be understood as the maximal Hubble 
rate during inflation.
Note that this bound does not depend on the 
statistical considerations developed in this section,  since it follows 
directly from the stability of the electroweak vacuum in our patch of the 
Universe. 

One could also envision the case in which there is a direct coupling between
the inflaton and the Higgs fields, say of the form $\phi^2h^2$. During 
inflation, this can play the role of a Higgs mass term, which can 
stabilize or destabilize the electroweak vacuum, depending on the sign of 
the corresponding coupling constant, in analogy with the case of the 
Higgs-curvature interaction.

\section{Implications for gravity wave signals}
\noindent
The considerations presented in the previous sections can have direct observational consequences once we realize
that the Hubble rate 
parametrizes 
the amount of tensor perturbations during inflation. During
the inflationary epoch, tensor perturbations, as for any other massless 
scalar
field, are quantum-mechanically generated. They can give rise to $B$-modes
of polarization of the CMB radiation
through Thomson scatterings of the CMB photons off free electrons
at last scattering \cite{pol}. The amplitude of the $B$-modes
depends on the amplitude of the gravity waves
generated during inflation, which in turn depends on the
energy scale at which inflation occured. The tensor-to-scalar power ratio
is often defined as $T/S$ and is given by $T/S \simeq (H_*/6.6\times 10^{13}\,
{\rm GeV})^2$. Current CMB anisotropy data impose the upper bound
$T/S\lsim 0.6$ \cite{Spergel:2006hy,wmapping}.

The possibility of detecting gravity waves from inflation via
$B$-modes is currently being considered by a number of ground, balloon and
space based experiments. The decomposition of the CMB polarization into
$E$- and $B$-modes requires a full sky data coverage and, as such, is limited
by the foreground contaminations. The latter introduce a mixing of the
$E$ polarization into   $B$ with the corresponding cosmic variance limitation.
Achieving levels below $T/S=10^{-4}$, or $H_*\lsim 
7\times 10^{11}$ GeV, requires observing 70 \% of the sky
with dust emission at 0.01 \% level \cite{sel}\footnote{  
Gravitational lensing also contaminates the tensor signal by converting the 
dominant $E$ polarization into $B$ polarization. Cleaning this 
contamination by reconstructing the lensing potential 
from CMB itself, 
one can achieve values of $T/S$ as small as $10^{-6}$ 
\cite{seljak}.}. 

\begin{figure} 
\label{gw}
\center{ \includegraphics[width=12cm,height=9cm]{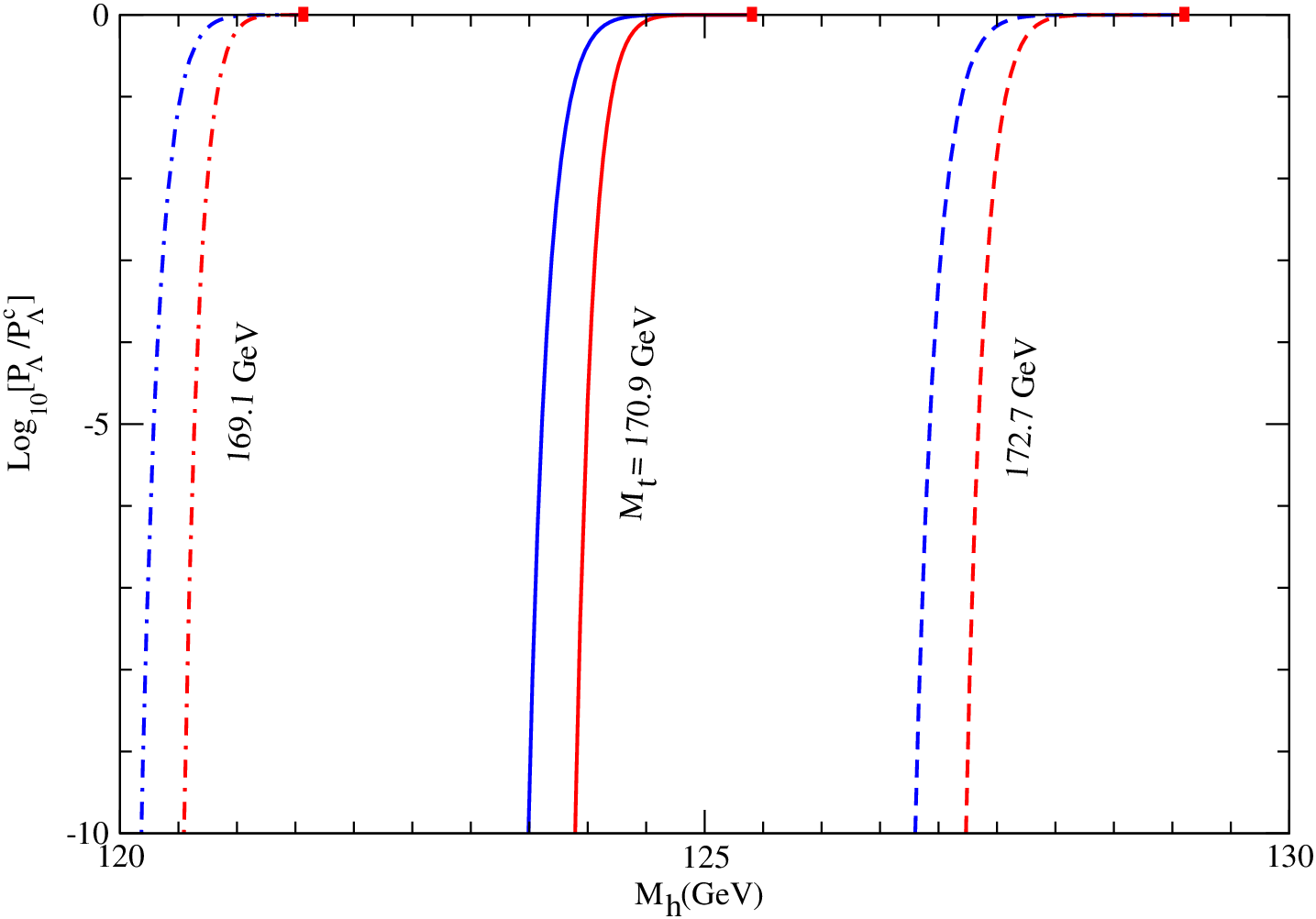}}
\caption{\label{fig:gw} The  probability 
$P_\Lambda$ versus the Higgs mass normalized with  its
value at $M_h^c$ for the Hubble rate $H_*=10^{13}$ GeV (upper curves) or
 $H_*=5\times 10^{13}$ GeV (lower curves), $N=N_*=60$ 
and for three
different values of the top mass for a model of inflation with
constant Hubble rate.} 
\end{figure}

Detection of the tensor mode would therefore imply that the value of the Hubble rate during inflation is  
larger than
about  $10^{12}$ GeV. This will have important implications for the
Higgs mass within all  inflationary models. In the  
vast  class of models where the potential is dominated by a constant
vacuum energy and the Hubble rate $H_*$ during inflation is roughly constant
(including the small-field and the hybrid models of inflation) 
the regions which survived the Higgs fluctuations are simply distributed
as
\begin{equation}
\label{kk}
P_\Lambda\propto e^{-\frac{H_*^2 N}{32\Lambda^2}}.
\end{equation}
This probability is dominated by the smallest value of the number of 
e-folds
and by the smallest value of the Hubble rate. Along the lines of the previous
section, we conclude that a successful detection of gravity waves with a value of $H_*$ larger than about
$10^{12}$ GeV, in conjunction with the discovery of a light Higgs, will indicate that we live in a highly improbable 
Universe. In fig.~\ref{gw}  we plot
the probability $P_\Lambda$ versus the Higgs mass normalized with  its
value at $M_h=M_h^c$ for the Hubble rate $H_*=10^{13}$ GeV (upper curves), 
$N=N_*=60$ 
and for three
different values of the top mass. We remind the reader that $M_h^c$ is the 
upper metastability limit of the Higgs mass given by  
\eq{stability} and shown by the upper curves in fig.~2. 

Similar results are also valid in the case of large-field models of inflation, with the only modification that $P_\Lambda$ is given by \eq{ll}. Therefore, in fig.~\ref{gw}
we also plot the probability $P_\Lambda$ versus the Higgs mass normalized 
with  its
value at $M_h=M_h^c$ for the Hubble rate $H_*=5\times 10^{13}$ GeV 
(lower curves), 
$N=N_*=60$ 
and for three
different values of the top mass. This value of $H_*$ corresponds to the Hubble rate necessary to reproduce the observed curvature perturbations for $p=2$, see \eq{phi}.

In conclusion, our Universe becomes exponentially unlikely if measurements of the Higgs mass and of the tensor-to-scalar power ratio $T/S$ violate the 
condition
\beq
\frac TS < \left( \frac{\Lambda}{10^{13}~{\rm GeV}}\right)^2,
\label{trul}
\eeq
where the value of $\Lambda$ has to be inferred from the measured value of the Higgs mass  from  fig.~1. 
The violation of \eq{trul} could then be interpreted as 
an indication for the existence of new physics beyond the SM, which affects the extrapolation of the Higgs potential to large energy scales.

\section{Conclusions}

In this paper we have investigated some possible cosmological implications
of the Higgs mass measurement. If the LHC discovers a Higgs with mass in
the metastability window shown in fig.~\ref{fig:window}, and does not find direct
evidence for other new physics, then there is a concrete possibility that we live in a metastable state. 
If we assume that the pure SM is
valid up to very large energy scales, then stability against field
fluctuations tests properties of the early Universe. We have revisited the
known considerations about thermal fluctuations, interpreting the result
as an upper limit on the reheating temperature $T_{RH}$ after inflation.
The bound is summarized in fig. \ref{mhTRH}.

We have also discussed the possibility  that the inflationary vacuum
fluctuations  destabilize the electroweak vacuum. 
If the Hubble rate is large enough during inflation and the Higgs mass
is light, then the danger exists that the  classical
value of the Higgs field is pushed above its instability point causing 
the collapse of the corresponding inflating domain. 
This does not necessarily pose a problem, since all the surviving domains
will be characterized by the correct electroweak vacuum. However, interesting
probabilistic conclusions can be reached under the assumption of a ``landscape" scenario in which
the inflationary parameters take different values in different patches of the Universe.
In the context of large-field models of inflation
we have argued that,  among 
the surviving regions, 
only an exponentially  tiny fraction of them  will be characterized 
by the correct amount of cosmological perturbations. 
If the Higgs mass is found below $(120-130)$~GeV, either we live in a very special, and exponentially unlikely, domain or new physics must exist below the scale $\Lambda$.

Moreover, our considerations
can be directly related to the amount of primordial gravity waves which can be
measured through their imprint on the CMB. The discoveries of a light Higgs boson and of tensor modes would be mutually conflicting (at least in a probabilistic sense) if the condition in \eq{trul} is not satisfied. Again, this evidence could be interpreted as an indication of new physics modifying the extrapolation of the Higgs potential to large field values. This result is valid, independently of the particular inflationary model considered.

Finally, a measurement of the Higgs mass below about 130~GeV provides a direct bound, shown in fig.~\ref{conformal}, on the quantity $\xi H^2$,
where $\xi$ is the (negative) coupling between the Higgs bilinear and the 
curvature, and $H$ is the maximal Hubble rate during inflation. This bound 
becomes particularly interesting in case of detection of primordial 
gravity waves, which provide a measurement of $H$. Note that the bound in 
fig.~\ref{conformal} does not depend on any statistical consideration of 
parameter scanning.

\section*{Acknowledgments}
J.R.E. thanks CERN for hospitality and partial financial support  
during the initial stages of this work. Work supported in part by CICYT, 
Spain, under
contracts FPA2004-02015; by a Comunidad de Madrid project (P-ESP-00346); 
and by the European Commission under contracts MRTN-CT-2004-503369 and 
MRTN-CT-2006-035863. 

\section*{Appendix}

In this appendix we present the relations between running couplings and
pole masses for the top quark and the Higgs, and the RG equations used in
our numerical calculation. We work at the next-to-leading order and use
the $\overline {\rm MS}$ renormalization scheme. We define
\beq
\lambda (M_t)=\frac{M_h^2}{2 v^2} \left( 1+\Delta_h\right) ,
\label{mhlambda}
\eeq
\beq
h_t (M_t) = \frac{\sqrt{2}}{v} M_t \left( 1+\Delta_t \right) .
\label{mtht}
\eeq
The corrections to the top-quark mass are given by
\beq
\Delta_t=\delta_t^{QCD}+\delta_t^{W}+\delta_t^{QED}\ .
\eeq
In the QCD part we include up to two-loop corrections using 
\cite{mtqcd1,mtqcd2}
\beq
\delta_t^{QCD}=-\frac{4\alpha_s(M_t)}{3\pi}+(1.0414 
N_f-14.3323)\left[\frac{\alpha_s(M_t)}{\pi}\right]^2\ ,
\eeq
where $N_f=5$. Although the two-loop effect is beyond the precision of our
computation, we include it because it is known to be large. The 
two-loop mixed electroweak-QCD \cite{mtmixed} and the three-loop
QCD corrections are also known \cite{mtqcd3}, but give only a small 
effect.

The electroweak part, including subleading corrections, is well 
approximated by \cite{mtew}
\bea 
\delta_t^{QED}+\delta_t^{W}&=&-\frac{4\alpha(M_t)}{9\pi}+\frac{h_t^2}{32\pi^2}
\left[\frac{11}{2}-r+2r(2r-3)\ln(4r)-8r^2\left(\frac{1}{r}-1\right)^{3/2}
\arccos\sqrt{r}\right]\nonumber\\
&-&6.90\times 10^{-3}+1.73\times 10^{-3}\ln\frac{M_h}{300\gev}-5.82\times 
10^{-3}\ln\frac{M_t}{175\gev}\ ,
\eea
where $r\equiv M_h^2/(4M_t^2)$. The formula above is valid for $r<1$, the 
case of interest for us. For $r\geq 1$ one has to replace 
$(1/r-1)^{3/2}\arccos\sqrt{r}$ by $(1-1/r)^{3/2}{\rm arccosh}\sqrt{r}$. 
The 
sign 
of $6.90\times 10^{-3}$ corrects \cite{gamb} the misprint of ref.~\cite{mtew}.

For the relation (\ref{mhlambda}) between the Higgs pole mass and the 
quartic coupling $\lambda$ we use \cite{sz}
\beq
 \Delta_h = \frac{G_F}{\sqrt{2}}\frac{M_Z^2}{16 \pi^2}
\left[\, \xi f_1(\xi) + f_0(\xi)  + \xi^{-1} f_{-1}(\xi) \,\right], 
\eeq
where
\bea
f_1(\xi) & = & 6\ln\frac{M_t^2}{M_h^2} +\frac{3}{2}\ln\xi
                     -\frac{1}{2} Z\!\left(\frac{1}{\xi}\right)
                     -Z\!\left(\frac{c_w^2}{\xi}\right)
                     -\ln c_w^2
                     +\frac{9}{2}
                        \left( \frac{25}{9} - 
\frac{\pi}{\sqrt{3}}\:\right),\\
 f_0(\xi) & = & -6\ln\frac{M_t^2}{M_Z^2}
                        \left[ 1 +2 c_w^2 -2\frac{M_t^2}{M_Z^2} \right]
                     +\frac{3 c_w^2 \xi}{\xi-c_w^2} \ln\frac{\xi}{c_w^2}
                     +2 Z\!\left( \frac{1}{\xi} \right) \nonumber\\
               &   & +\;4 c_w^2 Z\!\left( \frac{c_w^2}{\xi} \right)
                     +\left(\frac{3 c_w^2}{s_w^2} +12 c_w^2\right) \ln c_w^2
                     -\frac{15}{2} \left( 1 +2 c_w^2 \right) \nonumber\\
               &   & -\;3\frac{M_t^2}{M_Z^2} \left[
                        2 Z\!\left( \frac{M_t^2}{M_Z^2 \xi} \right)
                        +4 \ln\frac{M_t^2}{M_Z^2} -5 \right], \\
 f_{-1}(\xi) &=& 6\ln\frac{M_t^2}{M_Z^2}
                        \left[ 1 +2 c_w^4 -4\frac{M_t^4}{M_Z^4} \right]
                     -6 Z\!\left( \frac{1}{\xi} \right)
                     -12 c_w^4 Z\!\left( \frac{c_w^2}{\xi} \right)
                     -12 c_w^4 \ln c_w^2 \nonumber\\
               &   & +\;8\left( 1 +2 c_w^4 \right)
                     +24 \frac{M_t^4}{M_Z^4} \left[
                        \ln\frac{M_t^2}{M_Z^2} -2
                       + Z\!\left( \frac{M_t^2}{M_Z^2 \xi} \right) 
\right],
\eea
with $\xi \equiv M_h^2/M_Z^2,\, s_w^2 = \sin^2 \theta_W,\,
  c_w^2 = \cos^2 \theta_W$ and $\theta_W$ the Weinberg angle.
In addition,
\hfill\hfill\parbox{9cm}{\begin{eqnarray*}
 Z(z) &=& \left\{ \begin{array}{l@{\quad}l}
                2 A \arctan(1/A) &
                        (z > 1/4 ) \\
                A \ln \left[ (1+A)/(1-A) \right] &
                        (z < 1/4 )\; ,
        \end{array} \right.
\end{eqnarray*}}\hfill
\parbox{6cm}{\begin{equation}
 A = \sqrt{ \left| 1 - 4 z \right| }\; .\hspace*{1cm}\mbox{}
\end{equation}}

We also collect here the 2-loop renormalization-group equations that 
describe the evolution of the SM couplings and Higgs wave-function 
renormalization at scales beyond $M_t$ 
\cite{RGEs}. We keep only the gauge couplings, the top Yukawa coupling and 
the scalar quartic coupling. The 2-loop RG equations for the gauge 
couplings $g_i=\{g',g,g_s\}$ are
\beq
\frac{d g_i}{dt}=\kappa g_i^3b_i +\kappa^2 g_i^3
\left(\sum_{j=1}^3B_{ij}g_j^2-d_i^t h_t^2 \right) ,
\label{gauger}
\eeq
where $t=\ln Q$, $Q$ is the renormalization scale, $\kappa=1/(16\pi^2)$ 
and
\beq
b=\left(41/6,-19/6,-7\right)\ , ~~~~~
B=\left(\begin{array}{ccc}
199/18 & 9/2  & 44/3\\
3/2    & 35/6 & 12  \\
11/6   & 9/2  & -26
\end{array}\right)\ , ~~~~~
d^t=\left(17/6,3/2,2\right) \ .
\eeq

For the top Yukawa coupling
\beq
\frac{d h_t}{dt}=\kappa h_t\left( \frac{9}{2}h_t^2
-\sum_{i=1}^3c_i^tg_i^2\right)
+\kappa^2 h_t\left[\sum_{i j}D_{ij}g_i^2g_j^2+\sum_i
E_i g_i^2 h_t^2 + 6(\lambda^2 - 2 h_t^4 - 2 \lambda h_t^2) 
\right]\ ,
\label{yukk}
\eeq
with
\beq
c^t=\left(17/12, 9/4, 8\right)\ , ~~~~~
D=\left(\begin{array}{ccc}
1187/216   &   0   &   0  \\
 -3/4      & -23/4 &   0  \\
 19/9      &   9   & -108
\end{array}\right)\ , ~~~~~
E=\left(131/16,225/16,36\right)\ .
\eeq

The RG equation for the Higgs quartic coupling is
\bea
\frac{d\lambda}{dt} &=& \kappa\left\{-6h_t^4+12 h_t^2\lambda
+\frac{3}{8}\left[2g^4+(g^2+{g'}^2)^2\right]-3\lambda(3g^2+{g'}^2)
+24\lambda^2\right\}
\nonumber \\
&+&\kappa^2\left\{30 h_t^6-h_t^4\left(32 g_s^2 
+\frac{8}{3}{g'}^2+3\lambda\right)+h_t^2\left[-\frac{9}{4}g^4+\frac{21}{2} 
g^2 {g'}^2 -\frac{19}{4}{g'}^4\right.\right.\nonumber\\
&+&\left.\lambda\left(80g_s^2+\frac{45}{2}g^2+\frac{85}{6}{g'}^2
-144\lambda\right)\right]
+\frac{1}{48}\left(915g^6-289g^4{g'}^2-559g^2{g'}^4-379{g'}^6\right)
\nonumber\\
&+&\left.\lambda \left(-\frac{73}{8}g^4+\frac{39}{4}g^2{g'}^2
+\frac{629}{24}{g'}^4+108\lambda g^2 + 36 \lambda {g'}^2 -312\lambda^2
\right)\right\}\ .
\eea

Finally, the RG equation for the Higgs field wave function renormalization 
is 
\bea
\frac{1}{h}\frac{d h}{dt} &=& \kappa\left[-3 
h_t^2+\frac{3}{4}\left(3g^2 +{g'}^2\right)
\right]\\
&+&\kappa^2\left[\frac{27}{4}h_t^4-\frac{5}{2}h_t^2\left(8g_s^2
+\frac{9}{4}g^2+\frac{17}{12}{g'}^2\right)+\frac{271}{32}g^4-
\frac{9}{16}g^2{g'}^2-\frac{431}{96}{g'}^4-6\lambda^2
\right] \ .\nonumber
\eea

\end{document}